\documentclass[11pt]{article}

\usepackage[utf8]{inputenc}
\usepackage{cite}  
\usepackage{graphicx}
\usepackage[margin=1.25in]{geometry}
\usepackage[usenames,dvipsnames]{color}
\usepackage{url}
\usepackage[colorlinks = true,
            linkcolor = blue,
            urlcolor  = blue,
            citecolor = blue,
            anchorcolor = blue]{hyperref}
            
\usepackage{orcidlink}
\usepackage{authblk}

\usepackage{lineno}

\newcommand\pubnumber{v1}
\newcommand\pubdate{\today}

\def\Title#1{\begin{center} {\LARGE #1 } \end{center}}

\newcommand\pubblock{\rightline{\begin{tabular}{l} \pubnumber\\
         \pubdate \end{tabular}}}
\newenvironment{Abstract}{\begin{quotation} \begin{center}
                       ABSTRACT
     \end{center}\bigskip  }{\end{quotation}}

\newcommand\snowmass{\begin{center}\rule[-0.2in]{\hsize}{0.01in}\\\rule{\hsize}{0.01in}\\
\vskip 0.1in Submitted to the  Proceedings of the US Community Study\\ 
on the Future of Particle Physics (Snowmass 2021)\\ 
\rule{\hsize}{0.01in}\\\rule[+0.2in]{\hsize}{0.01in} \end{center}}

\def\beq{\begin{equation}}
\def\eeq#1{\label{#1}\end{equation}}
\def\eeqn{\end{equation}}

\newenvironment{Eqnarray}%
   {\arraycolsep 0.14em\begin{eqnarray}}{\end{eqnarray}}
\def\beqa{\begin{Eqnarray}}
\def\eeqa#1{\label{#1}\end{Eqnarray}}
\def\eeqan{\end{Eqnarray}}

\let\bar=\overbar

\def\lsim{\mathrel{\raise.3ex\hbox{$<$\kern-.75em\lower1ex\hbox{$\sim$}}}}
\def\gsim{\mathrel{\raise.3ex\hbox{$>$\kern-.75em\lower1ex\hbox{$\sim$}}}}

\def\del{\partial}
\def\Dslash{\not{\hbox{\kern-4pt $D$}}}
\def\dslash{\not{\hbox{\kern-2pt $\del$}}}
\def\pslash{\not{\hbox{\kern-2pt $p$}}}
\def\ETmiss{\not{\hbox{\kern-4pt $E$}}_T}

\def\Dlr{\mathrel{\raise1.5ex\hbox{$\leftrightarrow$\kern-1em\lower1.5ex\hbox{$D$}}}}

\def\MSB{{\bar{M \kern -2pt S}}}
\def\msb{{\bar{\scriptsize M \kern -1pt S}}}

\def\drb{{\bar{\scriptsize D \kern -1pt R}}}

\begin{document}

\pubblock

\snowmass{}

\Title{\textbf{Data Preservation for Cosmology\\
\vspace{10pt}%
TF09: Astro-particle physics \& cosmology \\
\vspace{10pt}%
COMPF7 (Reinterpretation and long-term preservation of data and code)}}
\medskip


\begin{center}
Marcelo Alvarez\textsuperscript{1},
Stephen Bailey\textsuperscript{1},
Deborah Bard\textsuperscript{2},
Lisa Gerhardt\textsuperscript{2},
Julien Guy\textsuperscript{1},
St\'ephanie Juneau\textsuperscript{4},
Anthony Kremin\textsuperscript{1},
Brian Nord\textsuperscript{3},
David Schlegel\textsuperscript{1},
Laurie Stephey\textsuperscript{2},
Rollin Thomas\textsuperscript{2},
Benjamin Weaver\textsuperscript{4}
\end{center}

\begin{center}
\textbf{1}~Lawrence Berkeley National Laboratory, USA
\textbf{2}~National Energy Research Scientific Computing Center (NERSC)
\textbf{3}~Fermi National Accelerator Laboratory, USA
\textbf{4}~NSF's National Optical-Infrared Astronomy Research Laboratory, USA
\end{center}

Corresponding Author Email: StephenBailey@lbl.gov

\begin{Abstract}
\noindent
We describe the needs and opportunities for preserving cosmology datasets and simulations,
and facilitating their joint analysis beyond the lifetime of individual projects.
We recommend that DOE fund a new cosmology data archive center to coordinate this work
across the multiple DOE computing facilities.
\end{Abstract}

\clearpage



\section{Executive Summary}

\subsection{Findings}

\begin{enumerate}
    \item NASA, NSF, ESA, and CERN fund data archive centers dedicated to preserving and sharing data beyond the lifetime of individual projects.  DOE does not have an equivalent data archive center for its Cosmic Frontier experiments and simulations.
    \item Current archive centers are focused on making data available for download, with limited support for providing computational resources for in-situ analysis on those data.
\end{enumerate}

\subsection{Comments}

\begin{enumerate}
    \item Currently the preservation of datasets after the operations phase is largely implemented on a best-effort basis by the lead labs.  This ad-hoc arrangement puts long term data preservation at risk and hinders their joint analysis.
    \item As datasets and simulations grow, a “take out” model of datasets available for manual download from multiple uncoordinated data centers becomes unwieldy.  Future work needs to focus on co-locating data with computing, and automating the coordination between multiple data/compute centers.
\end{enumerate}

\subsection{Recommendations}

\begin{enumerate}
    \item DOE should fund a multi-site cosmology data archive center to preserve Cosmic Frontier datasets and simulations, and facilitate their joint analysis across different computing centers.
\end{enumerate}

\section{Introduction}

It is common within cosmology and astronomy to publicly release datasets and simulations
to promote their use beyond the lifetime and goals of the original projects that generated the data.
It is also common for analyses to combine data and simulations across multiple experiments to
achieve results that would not have been possible with single experiments alone,
e.g.~\cite{Planck_plus_LRG, LegacySurveys, eBOSS_plus_LS, Amon_LensingClustering}.

Publicly released data should follow the Findable, Accessible, Interoperable, and Reusable
(FAIR) principles for scientific data management~\cite{FAIR-paper}.  This requires
more work for the original project than simply putting the data on a website for download.
Additionally, curating and maintaining these data beyond the operations phase of a project
requires dedicated personnel and hardware resources.  Making these datasets public has
significant long term scientific value, but requires dedicated resources to enable that.

DOE has both large cosmology datasets and large computing centers, but so far there has been
relatively little coordination to promote the joint use of these data while levering these compute
resources across projects and across centers.
The National Academies Decadal Survey ``Pathways to Discovery in Astronomy and Astrophysics for the 2020s''
\cite{decadal2020} chapter 4 includes a recommendation
``NASA and the National Science Foundation should explore mechanisms to improve
coordination among U.S.~archive centers and to create a centralized nexus for
interacting with the international archive communities.''
DOE archive centers with cosmology datasets should be included as well.
There is an unrealized opportunity to better coordinate this work in the future, both
within DOE resources, and between DOE and other agencies, e.g.~the efforts towards
jointly processing Euclid+Rubin+Roman datasets\cite{EuclidRubinRomanJSP}.
This white paper lays out some of the broad issues and
opportunities, while purposefully not suggesting a specific technical solution.

\section{Existing Cosmology/Astronomy Data Archive Centers}

\subsection{Non-DOE cosmology-related data archive centers}

NASA funds multiple data archive centers, broadly organized by the
wavelengths of data that they serve:
High Energy Astrophysics Science Archive Research Center
(HEASARC)\footnote{\url{https://heasarc.gsfc.nasa.gov}}
for extreme ultraviolet, X-ray, and gamma-ray wavelengths;
Mikulski Archive for Space Telescopes
(MAST)\footnote{\url{https://archive.stsci.edu}}
for optical, ultraviolet, and near-infrared;
NASA/IPAC Infrared Science Archive
(IRSA)\footnote{\url{https://www.ipac.caltech.edu/project/irsa}}
for infrared and submillimeter;
and the NASA Exoplanet Science Institute
(NExScI)\footnote{\url{https://nexsci.caltech.edu}}
for Exoplanet Exploration Program missions.
Additionally, the NASA/IPAC Extragalactic Database
(NED)\footnote{\url{http://ned.ipac.caltech.edu}}
curates catalog-level data about astronomical objects.

NSF archives its optical and infrared astronomy data through the NOIRLab Astro Data
Archive\footnote{\url{https://astroarchive.noirlab.edu}} and operates the Astro Data
Lab\footnote{\url{https://datalab.noirlab.edu}} science platform for database queries
and analysis tools.
This center has also ingested
subsets of other surveys such as SDSS and Gaia to facilitate cross matching these
data to core NSF datasets.

The Sloan Digital Sky Survey (SDSS)\footnote{\url{https://www.sdss.org}}
self-hosts its yearly data releases, providing access at a variety of levels
ranging from file downloads to database access to web queries and data visualization.

The European Space Agency (ESA) hosts data archives for each of its missions
at the European Space Astronomy Center (ESAC) Science Data Center
(ESDC)\footnote{\url{https://www.cosmos.esa.int/web/esdc}}.
The Centre de Données astronomiques de Strasbourg (CDS)\footnote{\url{https://cds.u-strasbg.fr}}
also curates and distributes multiple astronomy datasets.


With the exception of SDSS, these data archive centers are funded to preserve, curate, and share
data beyond the lifetime and budget of individual experiments.
Although they are hosted at multiple sites, the NASA data centers in particular
inter-operate to promote the discovery and use of these data across the different centers.

\subsection{How DOE cosmology projects currently share data}

In contrast to NASA and NSF, DOE-funded cosmology projects do not have a centrally coordinated
(and funded!) method of sharing their data, especially beyond the operations phase of each project.

The Dark Energy Survey (DES) public data releases are hosted my multiple non-DOE
sites\footnote{\url{https://des.ncsa.illinois.edu/releases/dr2/dr2-access}};
BOSS and eBOSS are hosted through SDSS\footnote{\url{https://www.sdss.org/dr17/}};
and the DESI Legacy Imaging Surveys\footnote{\url{https://legacysurvey.org}} are hosted at
NERSC via using resources of the Cosmology Data Repository allocation\footnote{\url{https://portal.nersc.gov/cfs/cosmo/data/legacysurvey/dr9/}},
and also available through the NOIRLab Astro Data Archive (images) and Astro Data Lab (catalogs).
DESI also intends to share its future spectroscopic data releases via the Cosmology Data Repository.

For cosmology simulations, 
Hardware/Hybrid Accelerated Cosmology Code (HACC)\footnote{\url{https://cosmology.alcf.anl.gov/}}
and AbacusSummit\footnote{\url{https://abacusnbody.org}} use the
``Modern Research Data Portal'' design pattern\cite{MRDP} to facilitate efficient bulk downloads
of subsets of their simulations, but these are independently implemented and hosted.
The Rubin Observatory Legacy Survey of Spact and Time (LSST)
Dark Energy Science Collaboration (DESC) data portal\footnote{\url{https://data.lsstdesc.org}}
also uses this design pattern to host its data from NERSC, but is completely independent
of both the Cosmology Data Repository and the AbacusSummit portal, both of which are also at NERSC.


\section{Co-locating Data with Computing Resources}

In most cases, existing cosmology/astronomy data archive centers
focus on enabling searching and downloading subsets of data,
rather than providing computing resources to the community to analyze the data in-situ.
Notable exceptions are the Astro Data Lab which provides a Jupyter Notebook server
for exploring the data\cite{DataLabJupyter},
and SDSS SciServer\footnote{\url{https://www.sciserver.org}}
to perform server side analysis of SDSS data\cite{SDSS_SciServer}.

Although the AbacusSummit files are available at NERSC to anyone with an account
(e.g.~LSST DESC members),
this is not emphasized in the portal documentation.  The same may be true of DESC simulations
(also at NERSC) and HACC (at ANL) --- the data are publicly available for download, and already
available at a major computing facility, but the current access methods still emphasize download
rather than in-situ use, even for those who already have accounts and allocations at the centers
where the original data are hosted.
The DESI Legacy Imaging Surveys may be unique
in documenting both how to download the data for those who do not have NERSC account,
and how to access the files directly on disc for those who do (e.g.~DESC or CMB-S4 members,
whether or not they are DESI collaborators).

At the same time, the datasets and simulations are large enough, and the computing needs of
analyses are diverse enough, that it is not realistic to expect that any one center could host
all of the data and meet the needs of all of the cosmology users.  A future cosmology data archive
center should promote both intra-site work (e.g.~DESC members at NERSC accessing public DESI data at NERSC)
and intra-site work (e.g.~combining simulations from the Argonne Leadership Class Facility with
DESI and CMB-S4 data at NERSC, LSST data at SLAC, and external datasets such as NASA archives).

\section{The Role of the Virtual Observatory}

The Virtual Observatory (VO) is ``vision that astronomical datasets and other resources should work as a seamless whole''\footnote{\url{https://www.ivoa.net}}, with the
International Virtual Observatory Alliance (IVOA) defining standards to enable this interoperability.
Practical end-user uptake has been somewhat slow, but now NASA, ESA, and the NOIRLab Astro Data Lab archive
centers provide VO-compliant interfaces for a subset of the most commonly used APIs
(e.g.~VO Table Access, Simple Image Access, Simple Spectal Access).
These are accessible through the astropy-affiliated
pyVO\footnote{\url{https://pyvo.readthedocs.io}}
and astroquery\footnote{\url{https://astroquery.readthedocs.io}} packages
which greatly simplify their usage.  The Rubin Science Platform also plans to provide
VO-compliant API interfaces as one of the data access methods.

The VO interfaces are primarily useful for classic astronomy use cases, e.g.~discovering and accessing distributed heterogeneous datasets covering a small number of individual objects.
It is not well suited for accessing the entirety of multi-terabyte cosmology survey data,
and thus is not well matched for may DOE cosmology projects which need more direct access to larger
volumes of homogeneous survey data.
e.g.~VO interfaces could be useful to discover if any prior survey has
a host galaxy redshift for a  new supernova candidate, but it would not be practical for doing a multi-band
pixel-level joint fit of all DECam plus WISE data.

Although the pyVO and astroquery packages have significantly simplified end-user access to
Virtual Observatory data, implementing these APIs to share new data remains non-trivial.
A DOE cosmology data archive center could assist in implementing these APIs to share subsets of the
data with the broader scientific community, but it should not be the primary method by which these
data are shared within the DOE Cosmic Frontier community.

\section{Brief Case Studies}

\subsection{DESI Legacy Imaging Surveys}

As a recent example,
the DESI Legacy Imaging Surveys\footnote{\url{https://legacysurvey.org}} were originally
motivated for DESI target selection though their public data release has led to a
large variety of diverse publications\footnote{\url{https://www.legacysurvey.org/pubs/}}.
This project combined pixel-level imaging data from the Dark Energy Survey and DECaLS (DECam on the CTIO Blanco 4-m telescope), MzLS (Kitt Peak Mayall 4-m), BASS (Kitt Peak Bok), WISE (NASA satellite), and catalog-level data from 2MASS, GAIA, and Pan-STARRS.  These data were jointly fit at NERSC using
pixel-level data across multiple imaging bands, originally downloaded with custom scripts
from 4 different data archive centers.
Much of the data processing was I/O bound and would not have been viable if trying to
download on-the-fly from the original data archive centers, some of which only provided HTTP download
of individual files as a data transfer method.  Co-locating the data with the computing resources
necessary to jointly analyze them was critical to the success of the project.

\subsection{Access to Rubin LSST Data from DOE Computing Facilities}

As a future example,
the Rubin Legacy Survey of Space and Time (LSST) will generate many petabytes of data, served from their
primary computing center at the SLAC National Laboratory.  DOE cosmology researchers have access to large
computing facilities at other sites, e.g.~NERSC or the Argonne and Oakridge Leadership Class Facilities,
and those researchers may need to use those computing resources to analyze LSST data.
Effectively analyzing these data from one site using computing resources at another site
will require either aggressive proactive caching, or non-trivial realtime data streaming on demand.
The scale of the data volume makes it unviable to perform bespoke downloads like those used by the
Legacy Imaging Surveys.

Additionally, some use cases such as DESC transient studies (e.g.~rapid followup of Type Ia supernovae),
will likely need nearly realtime access to other datasets such as DESI spectra of
host galaxies or photometry from imaging surveys preceding LSST, and those datasets may be originally
hosted at yet another data archive center.

Combining data at one site with computing at another site could be achieved through the efforts of individual collaborations (e.g.~DESC, CMB-S4, or future spectroscopic redshift surveys),
though it would be more effective to have a coordinated approach to solve the underlying challenges for everyone.

\section{Data distribution within industry}

Companies such as Google, Amazon, and Facebook automatically replicate their data across multiple data centers.
This is partially for robustness and data integrity so that all data are continuously available even
if there is an outage of an individual center, but it is also for performance so that commonly accessed
results (e.g.~images from trending topics) can be moved to fast storage at many centers, while infrequently
accessed data are automatically migrated to slower cheaper storage at fewer centers.

Within personal computing devices, it is common for photo and music libraries to be automatically synced
across multiple devices, including devices such as phones with limited storage that can't simultaneously hold
the entire dataset.  To the end user, this is transparent --- commonly accessed items are always there
(even when offline), and less frequently accessed items are automatically synced on demand as if they
always had been available.  Unlike current cosmology data management, the end user does not need to start
by freeing up storage space, identifying which data are at which locations, and initiating custom transfers
to get the data to a desired location before beginning analysis.

By analogy, one could imagine a multi-site cosmology data implementation where commonly used datasets are
at every computing center; less frequently accessed data are served by a primary host institution,
automatically replicated to another site for preservation robustness, and synced elsewhere as-needed upon demand;
and genuinely rarely accessed data are kept on tape archives.  Although all of this is currently possible
``by hand'', it lacks the automation and end-user transparency that would maximize the potential of using
large datasets across multiple large computing centers.

\section{Beyond cosmology datasets}

Although this white paper has focused on the specific needs of cosmology datasets and analyses,
the technical challenges are not unique to cosmology.  As such, this work could be done in the context
of a broader ``Experimental Data Archive Center'' to solve the more general problem of data preservation
and access, and thus leverage the resources of other communities such as Advanced Scientific Computing Research
(ASCR) and other areas within the DOE scientific portfolio beyond just High Energy Physics (HEP). 

\section{Conclusions}

DOE should fund a multi-site cosmology data archive center to preserve Cosmic Frontier datasets and simulations, and facilitate their joint analysis across different computing centers.  This requires support not only for hardware,
but also for personnel to develop and maintain the technologies to simplify cross-site data sharing
and personnel to curate the relevant datasets.

\def\thefootnote{\fnsymbol{footnote}}
\setcounter{footnote}{0}


\bibliographystyle{JHEP}
\bibliography{bib/preservation,bib/reinterpretation,bib/cosmo}

\providecommand{\href}[2]{#2}\begingroup\raggedright\begin{thebibliography}{10}

\bibitem{Planck_plus_LRG}
E.~{Kitanidis} and M.~{White}, \emph{{Cross-correlation of Planck CMB lensing
  with DESI-like LRGs}},
  \href{https://doi.org/10.1093/mnras/staa3927}{\emph{\mnras} {\bfseries 501}
  (2021) 6181} [\href{https://arxiv.org/abs/2010.04698}{{\ttfamily
  2010.04698}}].

\bibitem{LegacySurveys}
A.~{Dey}, D.J.~{Schlegel}, D.~{Lang}, R.~{Blum}, K.~{Burleigh}, X.~{Fan}
  et~al., \emph{{Overview of the DESI Legacy Imaging Surveys}},
  \href{https://doi.org/10.3847/1538-3881/ab089d}{\emph{\aj} {\bfseries 157}
  (2019) 168} [\href{https://arxiv.org/abs/1804.08657}{{\ttfamily
  1804.08657}}].

\bibitem{eBOSS_plus_LS}
P.~{Zarrouk}, M.~{Rezaie}, A.~{Raichoor}, A.J.~{Ross}, S.~{Alam}, R.~{Blum}
  et~al., \emph{{Baryon acoustic oscillations in the projected
  cross-correlation function between the eBOSS DR16 quasars and photometric
  galaxies from the DESI Legacy Imaging Surveys}},
  \href{https://doi.org/10.1093/mnras/stab298}{\emph{\mnras} {\bfseries 503}
  (2021) 2562} [\href{https://arxiv.org/abs/2009.02308}{{\ttfamily
  2009.02308}}].

\bibitem{Amon_LensingClustering}
A.~{Amon}, N.C.~{Robertson}, H.~{Miyatake}, C.~{Heymans}, M.~{White},
  J.~{DeRose} et~al., \emph{{Consistent lensing and clustering in a low-$S_8$
  Universe with BOSS, DES Year 3, HSC Year 1 and KiDS-1000}}, {\emph{arXiv
  e-prints} (2022) arXiv:2202.07440}
  [\href{https://arxiv.org/abs/2202.07440}{{\ttfamily 2202.07440}}].

\bibitem{FAIR-paper}
M.D.~Wilkinson et~al., \emph{{The FAIR Guiding Principles for scientific data
  management and stewardship}},
  \href{https://doi.org/10.1038/sdata.2016.18}{\emph{Scientific Data}
  {\bfseries 3} (2016) }.

\bibitem{decadal2020}
{National Academies of Sciences, Engineering, and Medicine}, \emph{{Pathways to
  Discovery in Astronomy and Astrophysics for the 2020s}} (2021),
  \href{https://doi.org/10.17226/26141}{10.17226/26141}.

\bibitem{EuclidRubinRomanJSP}
R.~{Chary}, G.~{Helou}, G.~{Brammer}, P.~{Capak}, A.~{Faisst}, D.~{Flynn}
  et~al., \emph{{Joint Survey Processing of Euclid, Rubin and Roman: Final
  Report}}, {\emph{arXiv e-prints} (2020) arXiv:2008.10663}
  [\href{https://arxiv.org/abs/2008.10663}{{\ttfamily 2008.10663}}].

\bibitem{MRDP}
K.~{Chard}, E.~{Dart}, I.~{Foster}, D.~{Shifflett}, T.~S. and J.~{Williams},
  \emph{{The Modern Research Data Portal: a design pattern for networked,
  data-intensive science}},
  \href{https://doi.org/https://doi.org/10.7717/peerj-cs.144}{\emph{PeerJ
  Computer Science 4:e144} (2018) }.

\bibitem{DataLabJupyter}
S.~{Juneau}, K.~{Olsen}, R.~{Nikutta}, A.~{Jacques} and S.~{Bailey},
  \emph{{Jupyter-Enabled Astrophysical Analysis Using Data-Proximate Computing
  Platforms}}, \href{https://doi.org/10.1109/MCSE.2021.3057097}{\emph{Computing
  in Science and Engineering} {\bfseries 23} (2021) 15}
  [\href{https://arxiv.org/abs/2104.06527}{{\ttfamily 2104.06527}}].

\bibitem{SDSS_SciServer}
M.~{Taghizadeh-Popp}, J.W.~{Kim}, G.~{Lemson}, D.~{Medvedev}, M.J.~{Raddick},
  A.S.~{Szalay} et~al., \emph{{SciServer: A science platform for astronomy and
  beyond}}, \href{https://doi.org/10.1016/j.ascom.2020.100412}{\emph{Astronomy
  and Computing} {\bfseries 33} (2020) 100412}
  [\href{https://arxiv.org/abs/2001.08619}{{\ttfamily 2001.08619}}].

\end{thebibliography}\endgroup

\end{document}